\documentclass[prb,twocolumn,amssymb,superscriptaddress]{revtex4-2}
\usepackage{amsmath}
\usepackage{amssymb}
\usepackage{amsthm}
\usepackage{amsfonts}
\usepackage[version=4]{mhchem}
\usepackage{listings}
\usepackage{enumerate}
\usepackage{latexsym}
\usepackage{psfrag}
\usepackage{bm}
\usepackage[all]{xy}
\usepackage{subfigure}
\usepackage[pdftex,colorlinks]{hyperref}
\usepackage{color}
\usepackage{mathtools}
\usepackage{times}
\usepackage{array}
\usepackage{placeins}

\begin{document}

\title{\ce{LnCu3(OH)6Cl3} (Ln = Gd, Tb, Dy): Heavy Lanthanides on Spin-1/2 Kagome Magnets}

\author{Ying Fu}
\affiliation{Joint Key Laboratory of the Ministry of Education, Institute of Applied Physics and Materials Engineering, University of Macau, Avenida da Universidade, Taipa, Macao SAR 999078, China}
\affiliation{Shenzhen Institute for Quantum Science and Engineering, and Department of Physics, Southern University of Science and Technology, Shenzhen 518055, China}

\author{Lianglong Huang}
\author{Xuefeng Zhou}
\author{Jian Chen}
\affiliation{Department of Physics, Southern University of Science and Technology, Shenzhen 518055, China}

\author{Xinyuan Zhang}
\affiliation{Institute of Functional Crystals, Tianjin University of Technology, Tianjin 300384, China}
\author{Pengyun Chen}
\affiliation{Institute of Resources Utilization and Rare-earth development, Guangdong Academy of Sciences, China}
\author{Shanmin Wang}
 \affiliation{Department of Physics, Southern University of Science and Technology, Shenzhen 518055, China}
\author{Cai Liu}
\affiliation{Shenzhen Institute for Quantum Science and Engineering, and Department of Physics, Southern University of Science and Technology, Shenzhen 518055, China}
\author{Dapeng Yu}
\affiliation{Shenzhen Institute for Quantum Science and Engineering, and
 	Department of Physics, Southern University of Science and Technology, Shenzhen
 	518055, China}

\author{Hai-Feng Li}
  \email{haifengli@um.edu.mo}
\affiliation{Joint Key Laboratory of the Ministry of Education, Institute of Applied Physics and Materials Engineering, University of Macau, Avenida da Universidade, Taipa, Macao SAR 999078, China}

  \author{Le Wang}
\email{wangl36@sustech.edu.cn}
\affiliation{Shenzhen Institute for Quantum Science and Engineering, and Department of Physics, Southern University of Science and Technology, Shenzhen 518055, China}
  \author{Jia-Wei Mei}
 \email{meijw@sustech.edu.cn}
 \affiliation{Shenzhen Institute for Quantum Science and Engineering, and
 	Department of Physics, Southern University of Science and Technology,
 	Shenzhen 518055, China}
 \affiliation{Shenzhen Key Laboratory of Advanced Quantum Functional Materials
 	and Devices, Southern University of Science and Technology, Shenzhen 518055, China}

\date{\today}

\begin{abstract}
The spin-1/2 kagome antiferromagnets are key prototype materials for studying frustrated magnetism. Three isostructural kagome antiferromagnets LnCu$_3$(OH)$_6$Cl$_3$ (Ln = Gd, Tb, Dy) have been successfully synthesized by the hydrothermal method. LnCu$_3$(OH)$_6$Cl$_3$ adopts space group $P\overline{3}m1$ and features the layered Cu-kagome lattice with lanthanide Ln$^{3+}$ cations sitting at the center of the hexagons. Although heavy lanthanides (Ln = Gd, Tb, Dy) in LnCu$_3$(OH)$_6$Cl$_3$ provide a large effective magnetic moment and ferromagnetic-like spin correlations compared to light-lanthanides (Nd, Sm, Eu) analogues, Cu-kagome holds an antiferromagnetically ordered state at around 17 K like YCu$_3$(OH)$_6$Cl$_3$.

\end{abstract}

\maketitle

\section{Introduction}

The kagome antiferromagnet (KAFM) has been intensively investigated both theoretically and experimentally as a long-standing platform to search for quantum spin liquid (QSL) \cite{Savary2016,Broholm2020,Feng2018a,Wen2019,Wei2020}, which is highly entangled quantum matter and features fractional excitations and no symmetry-breaking down to absolute zero temperature. Two famous KAFMs, Herbertsmithite (Cu$_3$Zn(OH)$_6$Cl$_2$) \cite{Helton2007,Han2012} and Zn-barlowite (Cu$_3$Zn(OH)$_6$FBr) \cite{Feng2017,Feng2018}, have been regarded as the prototype for QSL. Both of them show no phase transition down to low temperatures and exhibit fractional spinon excitations revealed by inelastic neutron scattering (INS) and nuclear magnetic resonance (NMR). Beyond QSL, additional interactions like Dzyaloshinskii-Moriya (DM) interactions and single-ion anisotropies may lead KAFM into other exotic ground states. For example, V$^{3+}$ (\emph{S} = 1) ions in NaV$_6$O$_{11}$ \cite{Seo1996,Kato2001} build kagome lattice and form spin-singlets. KFe$_3$(OH)$_6$(SO$_4$)$_2$ (\emph{S} = 5/2) \cite{Inami2000} presents a long-range order with positive chirality, while CdCu$_3$(OH)$_6$(NO$_3$)$_2$ (\emph{S} = 1/2) forms $120^{\circ}$ spin structure with negative chirality \cite{Nytko2009,Okuma2017}. Theoretically, a suitable combination of geometric frustration, ferromagnetism, and spin-orbit interactions in kagome magnets would realize high-temperature fractional quantum hall states, superconducting state \cite{Ribeiro2011,Guterding2016,Tang2011,Guo2009}. Experimentally, the Kondo physics scenario of non-magnetic impurities screened by spinons in QSL has been proposed according to the muon spin relaxation ($\mu$SR) study on ZnCu$_3$(OH)$_6$SO$_4$ \cite{Gomilsek2019}, analogous to Kondo effect usually observed in 3$d$-4$f$ heavy fermion metals, where local spins are screened by itinerant electrons.

Recently, YCu$_3$(OH)$_6$Cl$_3$ with perfect Cu-kagome layers and free of the Y-Cu anti-site disorder has been proposed as an ideal quantum KAFM \cite{Sun2016}, which has a ``$\mathbf{q} = 0$'' type (i.e., the magnetic unit cell is identical to structural unit cell with uniform chirality) antiferromagnetic (AFM) order with negative chirality due to a large DM interaction \cite{Zorko2019,Zorko2019a}. Replacing Yttrium with light lanthanides, RCu$_3$(OH)$_6$Cl$_3$ (R = Nd, Sm, Eu) compounds still show strongly frustrated behaviours in despite of forming the canted AFM order with Neel temperatures ($T_N$) ranging from 15 to 20 K \cite{Sun2017,Puphal2018}. With expecting that the heavy rare earths may further affect the magnetic frustration, we synthesized the polycrystalline samples of LnCu$_3$(OH)$_6$Cl$_3$ (Ln = Gd, Tb, Dy) by a universal way. The magnetic susceptibilities and heat capacity were measured. We discussed the magnetic contributions of Cu-kagome lattice and heavy lanthanides. With these results, we conclude that the heavy lanthanides ions in LnCu$_3$(OH)$_6$Cl$_3$ have little impact on the intrinsic magnetism of kagome-Cu$^{2+}$.

\section{Experimental}

\begin{figure*}[!t]
	\centering
	\includegraphics[width=1.8\columnwidth]{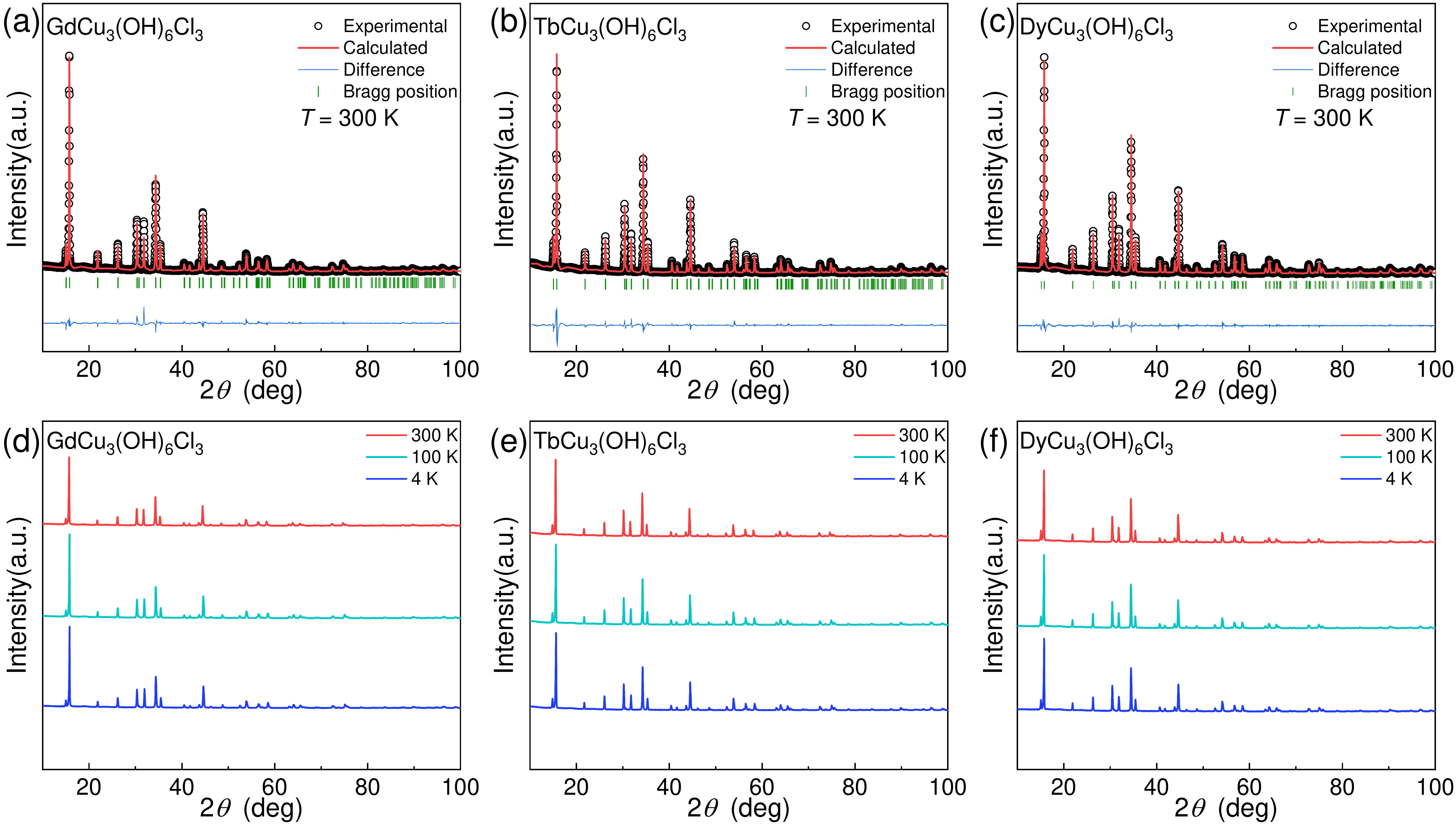}
	\caption{Powder X-ray diffraction patterns and refinements for LnCu$_3$(OH)$_6$Cl$_3$. (a)-(c) Refinements for PXRD at 300 K. (d)-(f) X-ray diffraction of LnCu$_3$(OH)$_6$Cl$_3$  at $T$ = 300, 100, 4 K.}
	\label{Fig1}
\end{figure*}

\begin{figure}[!t]
	\centering
	\includegraphics[width=0.9\columnwidth]{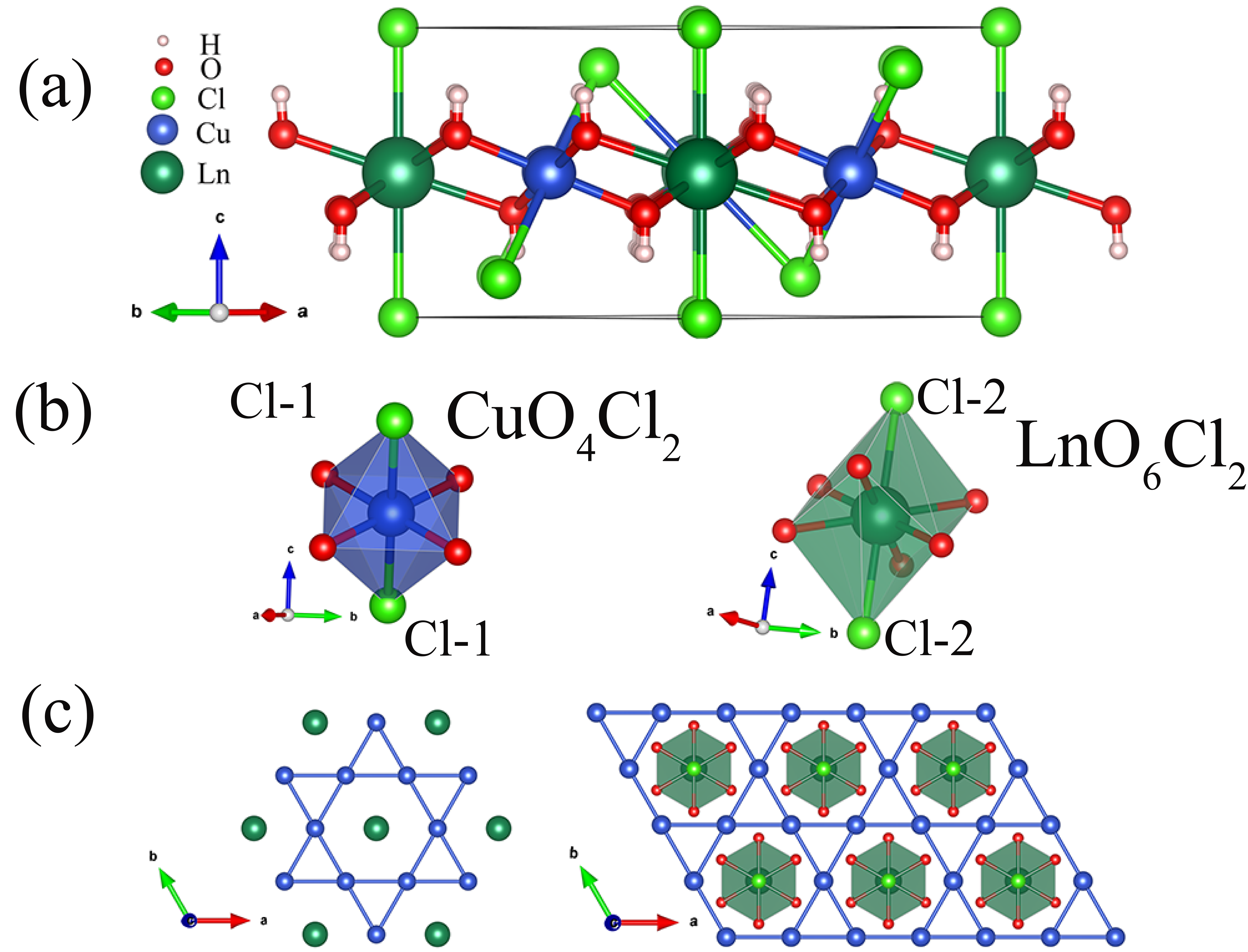}
	\caption{Crystal structure of LnCu$_3$(OH)$_6$Cl$_3$ (Ln = Gd,  Tb,  Dy) without considering the site-splitting of Ln$^{3+}$ ions. (a) Unit cell structure. (b) Coordinations of Cu and Ln atoms. (c) The illustration of Cu-kagome lattice and Ln-triangular lattice.}
	\label{Fig2}
\end{figure}

\begin{figure*}[!t]
	\centering
	\includegraphics[width=1.8\columnwidth]{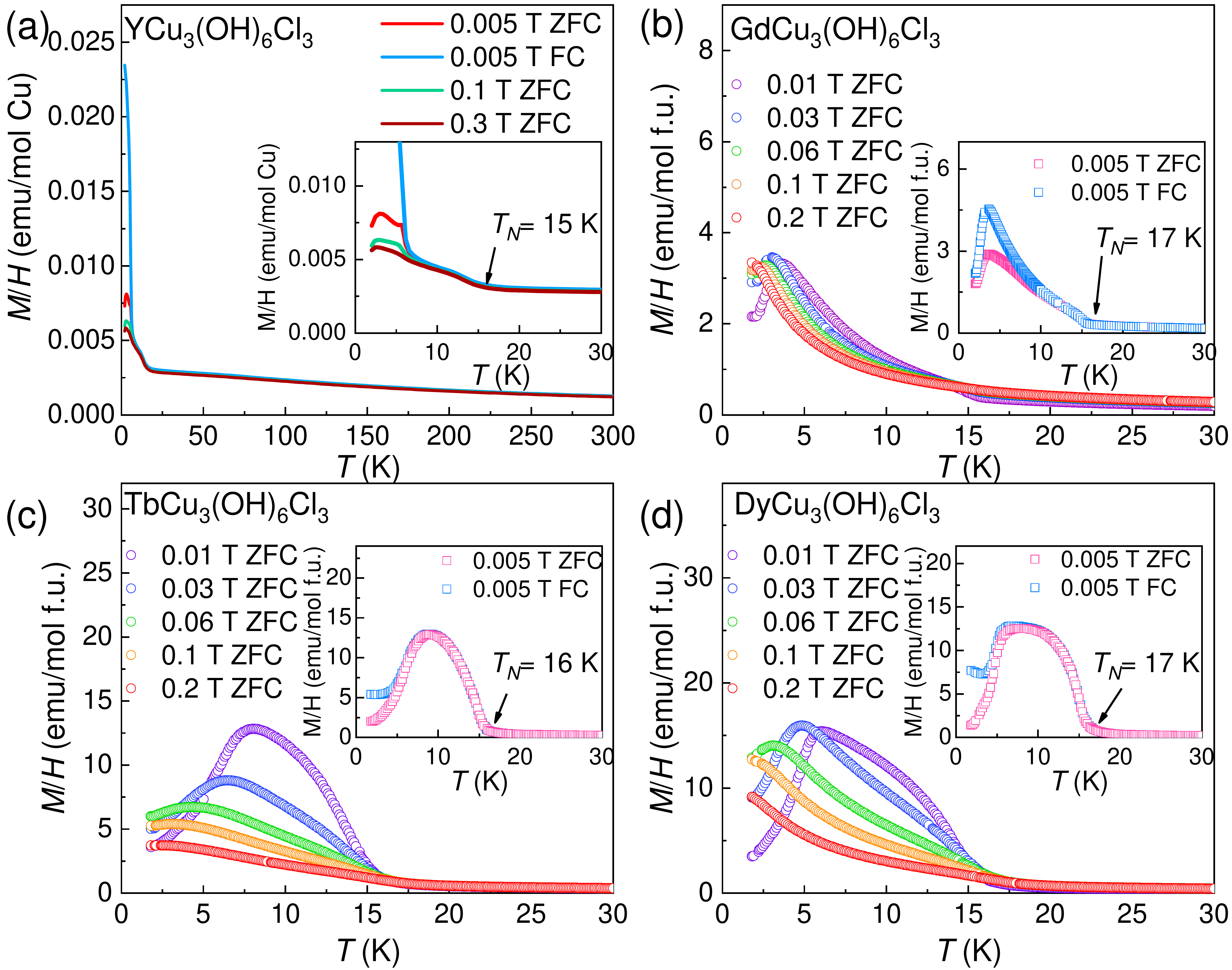}
	\caption{Temperature-dependent magnetization of YCu$_3$(OH)$_6$Cl$_3$ and LnCu$_3$(OH)$_6$Cl$_3$ (Ln = Gd, Tb, Dy) at selected fields. Inset in (a) is the zoom-in data, and insets in (b)-(d) are the ZFC and FC curves collected at 0.005 T.}
	\label{Fig3}
\end{figure*}

\begin{figure}[!t]
	\centering
	\includegraphics[width=0.9\columnwidth]{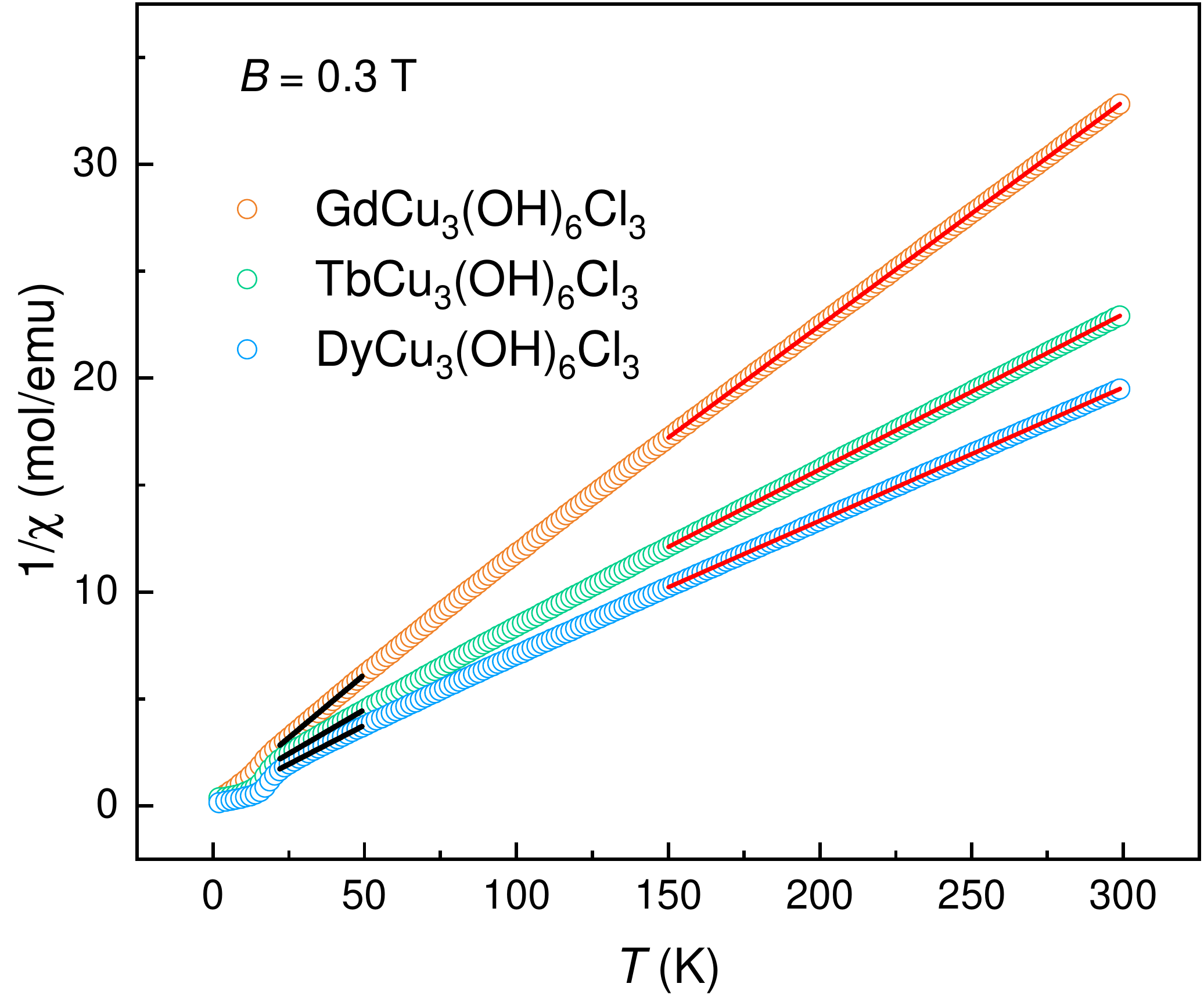}
	\caption{Temperature dependence of inverse magnetic susceptibilities $1/\chi$ under a field of 0.3 T. The red and black lines are the fitting-plots of Curie-Weiss law at high temperatures (150-300 K) and low temperatures (20-50 K), respectively.}
	\label{Fig4}
\end{figure}

\begin{figure*}[!t]
	\centering
	\includegraphics[width=1.8\columnwidth]{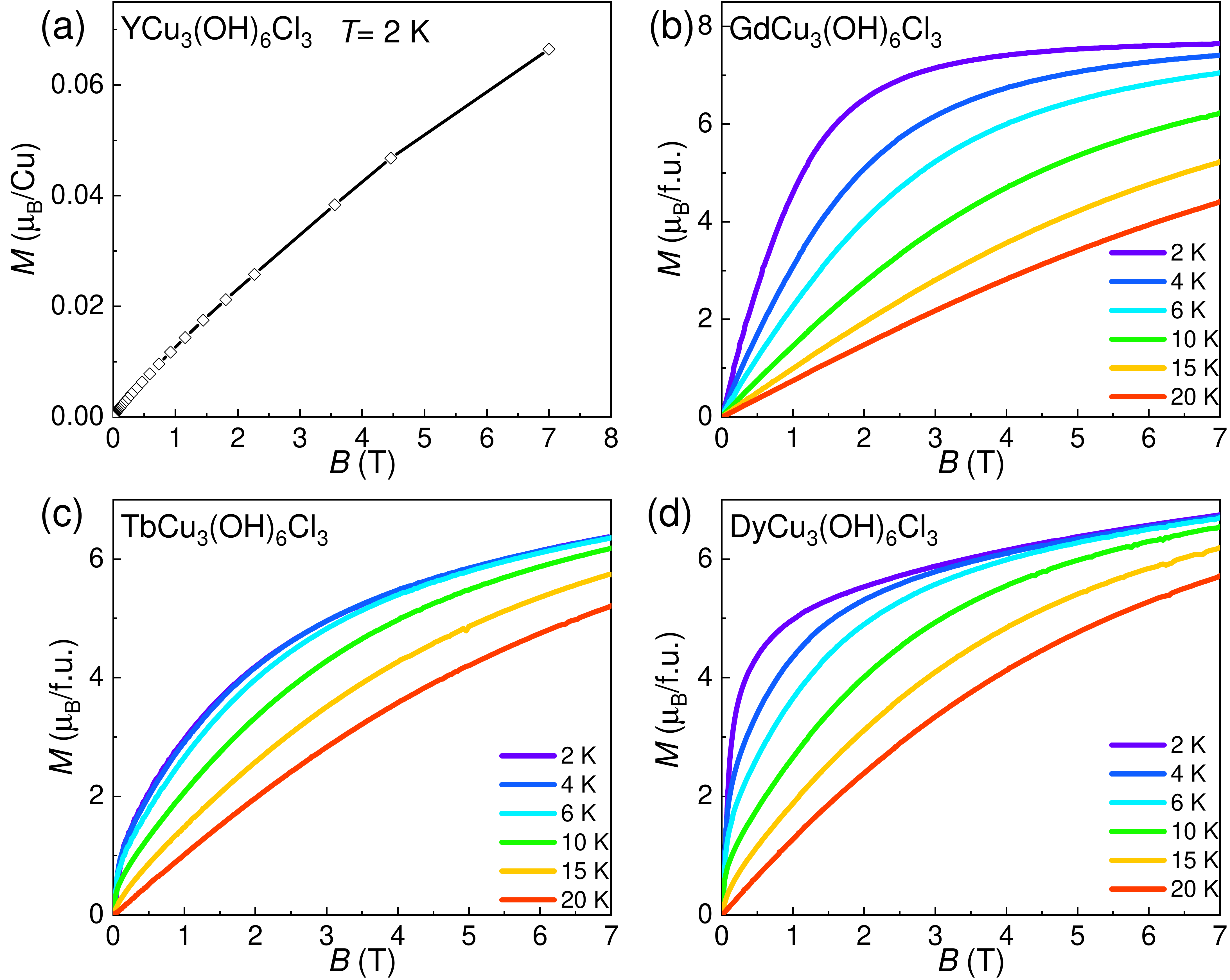}
	\caption{Field-dependent magnetization of (a) YCu$_3$(OH)$_6$Cl$_3$ at 2 K and (b)-(d) LnCu$_3$(OH)$_6$Cl$_3$ (Ln = Gd, Tb, Dy) at selected temperatures.}
	\label{Fig5}
\end{figure*}

\begin{figure*}[!t]
	\centering
	\includegraphics[width=1.8\columnwidth]{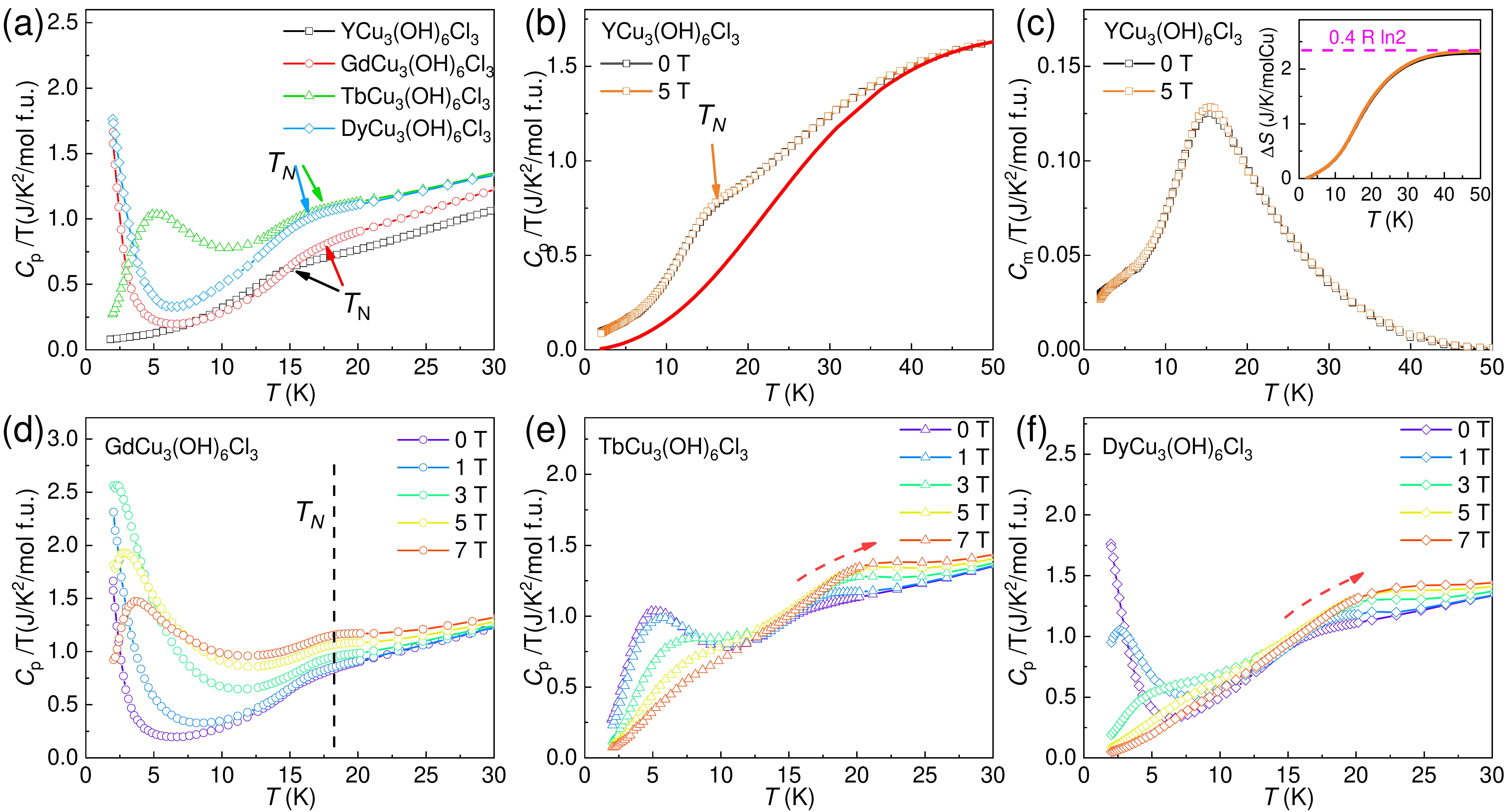}
	\caption{Specific heat for YCu$_3$(OH)$_6$Cl$_3$ and  LnCu$_3$(OH)$_6$Cl$_3$ (Ln = Gd, Tb, Dy). (a) $C_{p}/T $ for YCu$_3$(OH)$_6$Cl$_3$ and  LnCu$_3$(OH)$_6$Cl$_3$ under zero field. (b) The $C_p/T$ of YCu$_3$(OH)$_6$Cl$_3$. The red solid line is phonon-contribution fitting. (c) Magnetic specific heat  $C_m/T$ of YCu$_3$(OH)$_6$Cl$_3$ after subtracting phonon-contribution. Inset is magnetic entropy per Cu$^{2+}$. (d)-(f) Temperature-dependent specific heat under different magnetic fields for GdCu$_3$(OH)$_6$Cl$_3$, TbCu$_3$(OH)$_6$Cl$_3$ and DyCu$_3$(OH)$_6$Cl$_3$, respectively. }
	\label{Fig6}
\end{figure*}

Although the structure of GdCu$_3$(OH)$_6$Cl$_3$ was reported by Sun et al \cite{Sun2017} with tiny crystals, the high-purity sample was failed to obtain for further investigation of magnetic properties. In this work, we efficiently synthesized LnCu$_3$(OH)$_6$Cl$_3$ (Ln = Gd, Tb, Dy) samples with high purity by a hydrothermal method. The starting reagents were GdCl$_3$$\cdot$6H$_2$O (99.9\%, Alfa Aesar), TbCl$_3$$\cdot$6H$_2$O (99.99\%, Energy Chemical), DyCl$_3$$\cdot$6H$_2$O (99.99\%, Energy Chemical) and CuO (99.9\%, Alfa Aesar). LnCl$_3$$\cdot$6H$_2$O was ground thoroughly with CuO in a ratio of 1:3, and the mixture was transferred into an autoclave and heated at $200^{\circ}$C for about 10 h. Finally, the blue polycrystallined powder of LnCu$_3$(OH)$_6$Cl$_3$ was obtained after washed repeatedly by alcohol. This method avoids impurities and is also suitable for the preparation of YCu$_3$(OH)$_6$Cl$_3$, SmCu$_3$(OH)$_6$Cl$_3$ and EuCu$_3$(OH)$_6$Cl$_3$.

The temperature-dependent powder X-ray diffraction (PXRD) were performed from $T$ = 300 to 4 K on Rigaku Smartlab-9 kW diffractometer with Cu K$\alpha$ radiation ($\lambda_{K\alpha1 }$ = 1.54056 \AA, $\lambda_{K\alpha2}$ = 1.54439 \AA $ $, and intensity ratio $I_{K\alpha1}$ : $I_{K\alpha2}$ = 2:1). The scanning step width of $0.01^{\circ}$ was applied to record the patterns in a 2$\theta$ range of $10-100^{\circ}$. The structures of LnCu$_3$(OH)$_6$Cl$_3$ were refined by Rietveld profile methods using the FULLPROF Suite of programs \cite{RodriguezCarvajal1993}. The magnetic and specific heat measurements of LnCu$_3$(OH)$_6$Cl$_3$ were performed on  Quantum Design (QD) Magnetic Property Measurement System (MPMS) SQUID magnetometer and Physical Property Measurement System (PPMS), respectively.

\section{Results and Discussion}
\subsection{Crystal structure}

The site disorder remains mired in controversy in this series of compounds. For YCu$_3$(OH)$_6$Cl$_3$, single-crystal diffraction revealed a splitting disorder of Y$^{3+}$ and no anti-site disorder between Cu$^{2+}$ and Y$^{3+}$ \cite{Sun2016}. However, the neutron scattering results \cite{Barthelemy2019} supported no splitting disorder for Y$^{3+}$. The no site-splitting for rare earth ions was also proposed on LnCu$_3$(OH)$_6$Cl$_3$ (Ln = Nd, Sm, Gd, Eu) \cite{Sun2017,Puphal2018}. In this experiment, we found no obvious improvement for refinements after taking into account splitting disorder for rare earth ions. Therefore, we performed Rietveld profile refinement on LnCu$_3$(OH)$_6$Cl$_3$ (Ln = Gd, Tb and Dy) without considering the site-splitting of Ln$^{3+}$.

As shown in Figure \ref{Fig1}(a)-(c), the good refinements for LnCu$_3$(OH)$_6$Cl$_3$ in space group $P\overline{3}m1$ (no. 164) suggests that our powder sample is of high quality. The detailed lattice parameters are listed in Table \ref{table1}. As excepted at 300 K, the lattice parameters decrease from Gd$^{3+}$ to Dy$^{3+}$ in coincidence with the decreasing of ions radius ($r$ = 1.053 \AA, 1.04 \AA, 1.027 \AA $ $ for Gd$^{3+}$, Tb$^{3+}$ and Dy$^{3+}$, respectively). However, at 4 K, $a$ and $b$ show a contrast behaviour with $c$, which is associated with the anisotropic thermal expansion. The temperature-dependent XRD patterns, as shown in Fig.\ref{Fig1}(d)-(f), have no peak-splitting or new peaks appearing as temperature decreases down to 4 K, suggesting that no structure transition happens to LnCu$_3$(OH)$_6$Cl$_3$.

\begin{table}[b]
    \footnotesize
	\caption{\ Comparison of lattice parameters at 300 K and 4 K for LnCu$_3$(OH)$_6$Cl$_3$. LnCu$_3$(OH)$_6$Cl$_3$ is abbreviated to LnCu$_3$.}
	\label{table1}
	\begin{tabular*}{0.48\textwidth}{@{\extracolsep{\fill}}lll}
		\hline
		\hline
		& 300 K & 4 K \\
		\hline
		& \emph{a} = \emph{b } = 6.8019(47) \AA & \emph{a} = \emph{b} = 6.7923(65) \AA\\
		GdCu$_3$ & \emph{c} = 5.6240(12) \AA & \emph{c} = 5.5925(08) \AA\\
		& R$_{wp}$ = 9.47\%  & R$_{wp}$ = 10.0\%\\
		& R$_{Bragg}$= 3.76\% & R$_{Bragg}$= 4.21\%\\
		\hline		
		& \emph{a} = \emph{b} = 6.7888(35) \AA & \emph{a} = \emph{b} = 6.7797(91) \AA \\
		TbCu$_3$ & \emph{c} = 5.6219(72) \AA & \emph{c} = 5.5932(43) \AA \\
		& R$_{wp}$ = 13.1\% & R$_{wp}$  = 13.7\% \\
		& R$_{Bragg}$ = 6.35\% & R$_{Bragg}$ = 7.13\%\\
		\hline
		& \emph{a} = \emph{b} = 6.7655(98) \AA & \emph{a} = \emph{b} = 6.7631(14) \AA \\
		DyCu$_3$ &  \emph{c} = 5.6130(2) \AA & \emph{c} = 5.6052(97) \AA \\
		& R$_{wp}$ = 6.61\% & R$_{wp}$  = 6.61\% \\
		& R$_{Bragg}$ = 3.58\% & R$_{Bragg}$ = 3.9\% \\
		\hline
		\hline
	\end{tabular*}
\end{table}

As depicted in Figure \ref{Fig2}, each Cu$^{2+}$ is surrounded by four equivalent O$^{2-}$ and two Cl$^{-}$, forming a distorted [CuO$_4$Cl$_2$] octahedron with the Cu-Cl bond ($\sim$ 2.82 \AA) significantly longer than the Cu-O bond ($\sim$ 1.97 \AA). The [CuO$_4$Cl$_2$] octahedrons connect to each other by sharing the O-Cl edges to build the Cu-kagome plane. Ln$^{3+}$ is 8-coordinated by six O$^{2-}$ and two Cl$^{-}$ to form [LnO$_6$Cl$_2$] dodecahedron and locates the center of the Cu-hexagon to constitute Ln-triangular lattice (Fig.\ref{Fig2}(c)).

It is worth noting that analogous distorted octahedra [CuO$_4$Cl$_2$] in herbertsmithite and Y$_3$Cu$_9$(OH)$_{19}$Cl$_8$ increase the splitting in ligand-field of $d$ orbitals and lower the energy level of $d_{z^2}$ with a large $d$-$d$ gap around 1--2 eV, unveiling the insulating nature of a charge transfer insulator \cite{Pustogow2017}, which would  be adapted to YCu$_3$(OH)$_6$Cl$_3$ and LnCu$_3$(OH)$_6$Cl$_3$.

The Cu-O-Cu super-exchange bond angles are $118.78(12)^{\circ}$ for GdCu$_3$(OH)$_6$Cl$_3$, $118.19(10)^{\circ}$ for TbCu$_3$(OH)$_6$Cl$_3$, and $117.7(4)^{\circ}$ for DyCu$_3$(OH)$_6$Cl$_3$, respectively. The values are comparable to the antiferromagnets like barlowite ($117.4^{\circ}$) \cite{Han2014} and herbertsmithite ($119^{\circ}$) \cite{Freedman2010}. The Cu-Cu distances are 3.40330(11) \AA~for GdCu$_3$(OH)$_6$Cl$_3$, 3.39115(16) \AA~for TbCu$_3$(OH)$_6$Cl$_3$, and 3.3830(3) \AA~for DyCu$_3$(OH)$_6$Cl$_3$, equal to the corresponding Ln-Cu distance.

\subsection{Magnetic properties}
Figure \ref{Fig3} shows the temperature-dependent magnetizaion for LnCu$_3$(OH)$_6$Cl$_3$ (Ln = Gd, Tb, Dy), with YCu$_3$(OH)$_6$Cl$_3$ served as a reference. For YCu$_3$(OH)$_6$Cl$_3$ (Fig.~\ref{Fig3}(a)), the magnetic susceptibilities increase suddenly at 15 K, which was associated with the negative-vector-chirality 120$^{\circ}$ magnetic structure, confirmed by the neutron scattering study \cite{Zorko2019a}. Continuously lowering temperature, the drops at 3--4 K correspond to possible spin-glass state \cite{Zorko2019}. Compared to the small magnetic moment of Cu in YCu$_3$(OH)$_6$Cl$_3$, magnetic susceptibilities of LnCu$_3$(OH)$_6$Cl$_3$ are much larger, suggesting that the dominant contribution to the magnetization arises from lanthanides, especially at low temperatures. As shown in Fig.~\ref{Fig3}(b)-(d), under low fields, LnCu$_3$(OH)$_6$Cl$_3$ compounds present similar temperature-dependent magnetization curves as YCu$_3$(OH)$_6$Cl$_3$, with a rapid increase at about 16 or 17 K and followed by a drop at lower temperatures. The Zero field-cooling (ZFC) and field-cooling (FC) data at 0.005 T begin to split after the rapid increase, which could be ascribed to the possible in-plane canted ferromagnetic component, as previously pronounced in RCu$_3$(OH)$_6$Cl$_3$ (R = Nd, Sm, Eu) \cite{Sun2017,Puphal2018}. It indicates that Cu-kagome lattice of LnCu$_3$(OH)$_6$Cl$_3$ may have the same physics as YCu$_3$(OH)$_6$Cl$_3$, and form magnetic structure at $T_N \sim 16$ K, which necessitates a neutron scattering study. In contrast to the robust magnetic order at 15 K in YCu$_3$(OH)$_6$Cl$_3$, magnetic phase transition of LnCu$_3$(OH)$_6$Cl$_3$ can not be identified easily with increasing field. Whether $T_N$ was suppressed by fields or the magnetic responses of Ln$^{3+}$ ions masked the magnetic order of Cu$^{2+}$ needs a further measurement of specific heat.

As shown in Figure~\ref{Fig4}, the high temperature behaviour of inverse magnetic susceptibility above 150 K was fit to the Curie-Weiss law $\chi=C/(T-\theta)$ (where $C$ is the Curie constant, and $\theta$ is the Weiss temperature) in red lines with $C=$ 9.52 K$\cdot$emu$\cdot$mol$^{-1}$, 13.76 K$\cdot$emu$\cdot$mol$^{-1}$, 16.05 K$\cdot$emu$\cdot$mol$^{-1}$ and $\theta=$ $-$13.98 K, $-$16.37 K, $-$13.72 K for Gd-, Tb- and Dy-compounds, respectively. Considering the crystal-field splitting for Ln$^{3+}$ ions, we also applied the Curie-Weiss fitting between 20 K to 50 K, shown in black lines. The deduced $\theta$ are $-$1.71 K, $-$4.24 K, $-$1.40 K for GdCu$_3$(OH)$_6$Cl$_3$, TbCu$_3$(OH)$_6$Cl$_3$ and DyCu$_3$(OH)$_6$Cl$_3$, respectively. The absolute values of $\theta$ are smaller than that at high temperatures. The small Weiss temperature $\theta$ values are significant different from Nd-, Sm- and Eu-analogues \cite{Sun2017,Puphal2018}, whose $\theta$ ranges from $-100$ K to $-300$ K with a distinct spin frustration compared to $T_N$ $\sim$ 15 K. Two main reasons for the reduction of $\theta$ are proposed: One is that the Cu-Cu AFM-interaction is weakened by Ln$^{3+}$; another is that Ln$^{3+}$ ions form a FM-interaction that competes with the Cu-Cu AFM-interaction.

The effective magnetic moments $\mu_{eff}$ of LnCu$_3$(OH)$_6$Cl$_3$ deduced by the formula $\mu_{eff}=\sqrt{\frac{3k_BC}{N_A}}$ (where $k_B$ is the Boltzmann constant, $N_A$ is the Avogadro number, and $C$ is the Curie constant) are 8.21 $\mu_B$, 9.82 $\mu_B$ and 10.45 $\mu_B$ for GdCu$_3$(OH)$_6$Cl$_3$, TbCu$_3$(OH)$_6$Cl$_3$ and DyCu$_3$(OH)$_6$Cl$_3$, respectively. For LnCu$_3$(OH)$_6$Cl$_3$, a system consisting two kinds of magnetic ions Ln$^{3+}$ and Cu$^{2+}$ (the theoretical effective magnetic moment for single ions are 1.73 $\mu_B$  for Cu$^{2+}$, 7.94 $\mu_B$ for Gd$^{3+}$, 9.72 $\mu_B$  for Tb$^{3+}$ and 10.63 $\mu_B$ for Dy$^{3+}$), the total theoretical effective value of magnetic moment can be roughly estimated by formula $\mu_{eff}^{t}= \sqrt{\mu_{Ln^{3+}}^{2}+3\mu_{Cu^{2+}}^{2}}$, yielding $\mu_{GdCu_3}^{t}$ = 8.48 $\mu_B$, $\mu_{TbCu_3}^{t}$= 10.17 $\mu_B$, and $\mu_{DyCu_3}^{t}$ = 11.04 $\mu_B$. Unlike TbCu$_3$(OH)$_6$Cl$_3$ and DyCu$_3$(OH)$_6$Cl$_3$, the effective magnetic moment of GdCu$_3$(OH)$_6$Cl$_3$ is close to the theoretical value, suggesting that Gd$^{3+}$ is fully at the ground state due to the large crystal field-splitting energy $\Delta_{CEF}>>300$ K.

The field-dependent magnetization curves ($M$-$H$) of LnCu$_3$(OH)$_6$Cl$_3$ (Ln = Gd, Tb, Dy) are shown in Figure~\ref{Fig5}, referred to YCu$_3$(OH)$_6$Cl$_3$ collected at 2 K. With decreasing temperature, magnetization of LnCu$_3$(OH)$_6$Cl$_3$  increases and grows rapidly below a field of about 2 T. At 2 K, GdCu$_3$(OH)$_6$Cl$_3$ saturates to a large value of 7.69 $\mu_B$ at 7 T, suggesting that spins of Gd$^{3+}$ ions are polarized. TbCu$_3$(OH)$_6$Cl$_3$ and DyCu$_3$(OH)$_6$Cl$_3$ seem to be saturated with a linear increase under higher fields, especially for DyCu$_3$(OH)$_6$Cl$_3$, which could be ascribed to the temperature-independent Van Vleck paramagnetism. It is noted that the magnetic moment of each Cu$^{2+}$ at 2 K is only 0.065 $\mu_B$ in the strong frustrated material, YCu$_3$(OH)$_6$Cl$_3$ ($\mu_0H = $ 7 T) (Fig.~\ref{Fig5}(a)), which is two orders of magnitude smaller than that of LnCu$_3$(OH)$_6$Cl$_3$. Thus, we deduced that lanthanide has a dominated magnetic contribution in LnCu$_3$(OH)$_6$Cl$_3$ at low temperatures and can easily dominate the magnetic response of Cu-kagome lattice.

\subsection{Specific heat}

Figure \ref{Fig6} shows the specific heat results of LnCu$_3$(OH)$_6$Cl$_3$ (Ln = Gd, Tb, Dy) and YCu$_3$(OH)$_6$Cl$_3$. As shown in Fig.~\ref{Fig6}(a), under zero field, a shoulder anomaly was observed at around 15--17 K for each compound, consistent with the rapid increase of magnetic susceptibilities at $T_N$, representing a formation of magnetic order for kagome-Cu$^{2+}$. Moreover, the low temperature (below 10 K) specific heat of LnCu$_3$(OH)$_6$Cl$_3$ shows more features, in contrast to decaying to zero for YCu$_3$(OH)$_6$Cl$_3$, relating to the low-temperature magnetic correlation of Ln$^{3+}$ ions.

For further understanding the origin of magnetic phase transition, we measured the specific heat with applied magnetic fields. Since the intrinsic nearest neighboring interaction in YCu$_3$(OH)$_6$Cl$_3$ is around 80 K \cite{Arh2020}, the applied magnetic field (5 T) has little impact on specific heat and entropy (see Figs. \ref{Fig6}(b) and (c)), in line with previous report on YCu$_3$(OH)$_6$Cl$_3$ \cite{Zorko2019} and EuCu$_3$(OH)$_6$Cl$_3$ \cite{Puphal2018}. However, as shown in Figs. \ref{Fig6} (d)-(f), $C_p/T$ of LnCu$_3$(OH)$_6$Cl$_3$ responses notably to external magnetic field. For GdCu$_3$(OH)$_6$Cl$_3$, the upturn of $C_p/T$ is generally evolved into a broad peak and pushed to high temperatures by field with $T_N$ keeping constant. For TbCu$_3$(OH)$_6$Cl$_3$ and DyCu$_3$(OH)$_6$Cl$_3$, the low-temperature broad peak of $C_p/T$ is efficiently pushed to above $T_N$ by a field of 7 T and merges with the high-temperature broad peak induced by the magnetic phase transition. This behavior of driving peak position to high temperatures by applied magnetic fields in specific heat may indicate a formation of short-range ferromagnetic order below $T_N$.

Considering YCu$_3$(OH)$_6$Cl$_3$ forms a robust ``$\mathbf{q} = 0$'' type AFM order and RCu$_3$(OH)$_6$Cl$_3$ (R = Nd, Sm, Eu) enters a canted AFM phase below $T_N$, we speculated reasonably that LnCu$_3$(OH)$_6$Cl$_3$ (Ln = Gd, Tb, Dy) also happens a canted AFM phase transition at $T_N$ with a large ferromagnetic component. The ferromagnetic correlation is influenced obviously by fields and even screen the signal of AFM ordering in magnetic susceptibility, but the AFM phase indeed exists and is robust under large fields, like the case in YCu$_3$(OH)$_6$Cl$_3$.

\subsection{Discussion and Conclusions}

The $\mathbf{q} = 0$ type magnetic structure with negative-chirality in YCu$_3$(OH)$_6$Cl$_3$ is interesting, which was also reported in other kapellasite-type compounds like CdCu$_3$(OH)$_6$(NO$_3$)$_2$ with $T_N$ = 4 K \cite{Okuma2017} and CaCu$_3$(OH)$_6$Cl$_2$ with $T_N$ = 7.2 K \cite{Yoshida2017,Iida2020}. As demonstrated recently, with light lanthanides (Sm and Eu) replacing Yttrium, SmCu$_3$(OH)$_6$Cl$_3$ and EuCu$_3$(OH)$_6$Cl$_3$ still feature canted antiferromagnetic ordering with strong spin frustration \cite{Sun2017,Puphal2018}. The light lanthanides with small magnetic moment may have limited influence on magnetism of Cu-kagome lattice.

In our work, the magnetic and thermodynamic behaviors of LnCu$_3$(OH)$_6$Cl$_3$ (Ln = Gd, Tb, Dy) exhibit two significantly different characteristics: large magnetic moment compared with YCu$_3$(OH)$_6$Cl$_3$ and a ferromagnetic-like spin correlation below $T_N$. According to our experimental results, heavy lanthanides (Gd, Tb, Dy) probably modulate the DM interaction and induce a large ferromagnetic correlation, which can mask the intrinsic low-temperature magnetic properties of kagome-Cu$^{2+}$, but can not prevent the AFM ordering of Cu-kagome as revealed in specific heat. The Curie-Weiss law no longer works for evaluating the intrinsic interactions. The spectroscopy technology, like electron spin resonance (ESR) or $\mu$SR , is hopeful to further detect the detailed magnetic interactions for LnCu$_3$(OH)$_6$Cl$_3$ (Ln = Nd, Sm, Eu, Gd, Tb, Dy).

In summary, we have successfully synthesized the polycrystalline samples of LnCu$_3$(OH)$_6$Cl$_3$ (Ln = Gd, Tb and Dy). The heavy lanthanides significantly change the magnetic and thermodynamic behaviors, which keeps the intrinsic magnetism of Cu-kagome lattice. LnCu$_3$(OH)$_6$Cl$_3$(Ln = Nd, Sm, Eu, Gd, Tb, Dy) compounds provide a good platform to further investigate systemically the effect of lanthanides on the frustrated magnetism of Cu-kagome lattice.

\section*{Conflicts of interest}
There are no conflicts to declare.

\section*{Acknowledgements}
We thank Dr. L. Zhang, Dr. J. M. Sheng and Prof. L. S. Wu for useful discussion. This work was supported by the program for Guangdong Introducing Innovative and Entrepreneurial Teams (No. 2017ZT07C062), Shenzhen Key Laboratory of Advanced Quantum Functional Materials and Devices (NO. ZDSYS20190902092905285) and Guangdong Basic and Applied Basic Research Foundation
(No. 2020B1515120100). L. Wang acknowledges the support of China Postdoctoral Science Foundation (2020M682780). H.-F. Li acknowledges the financial supports from Science and Technology Development Fund, Macao SAR (File No. 0051/2019/AFJ), Guangdong Basic and Applied Basic Research Foundation (Guangdong-Dongguan Joint Fund No. 2020B1515120025), and Guangdong-Hong Kong-Macao Joint Laboratory for Neutron Scattering Science and Technology (Grant No. 2019B121205003).

\bibliographystyle{iopart-num.bst}
\bibliography{refs}

\end{document}